\documentclass[12p]{iopart}

\usepackage{graphicx}
\usepackage{iopams}

\begin{document}

\title[Numerical viscosity in GR hydrodynamics]{Numerical viscosity in hydrodynamics simulations
in general relativity}

\author{P. Cerd\'a-Dur\'an}
\address{Max-Planck-Institut f\"ur Astrophysik,
Karl-Schwarzschild-st. 1, D-85741 Garching bei M\"unchen, Germany}
\ead{cerda@mpa-garching.mpg.de}

\begin{abstract}
We present an alternative method to estimate the numerical viscosity in simulations
of astrophysical objects, which is based in the damping of fluid oscillations.
We apply the method to general relativistic hydrodynamic simulations using
spherical coordinates. We perform 1D-spherical and 2D-axisymmetric
simulations of radial oscillations in spherical systems. We calibrate first
the method with simulations with physical bulk viscosity and study the differences between
several numerical schemes. We apply the method to radial oscillations
of neutron stars and we conclude that the main source of numerical viscosity in this case
is the surface of the star. We expect that this method could
be useful to compute the resolution requirements and limitations of the numerical simulations
in different astrophysical scenarios in the future.
\end{abstract}


\submitto{\CQG}


\section{Introduction}

There is a number of scenarios in theoretical astrophysics which need to be modelled
by means of numerical simulations. The complexity found in these scenarios does not allow for
a simple analytical study to explain the observations. In a subset of these scenarios the
effects of general relativity are important and a Newtonian description of the system is not 
sufficient. Some examples of stellar size scenarios of this kind are: i) the collapse of massive 
stellar cores to form a neutron star, which eventually can lead to a supernova explosion,
ii) for more massive stars, the collapse to a black hole and possibly an accretion 
disk, which can power a long-duration gamma-ray burst (GRB) according to the collapsar scenario
(Woosley 1993) and could be related to the subclass of broad-lined type Ic supernovae (Modjaz \etal 2008),
and iii) the cooling of a rapidly rotating hot proto-neutron star (PNS) to form a
magnetar (Thomson \& Duncan 1993). All three previous examples have in common
that are quasi-spherical scenarios. Therefore it is natural to model them in
numerical simulations using spherical polar coordinates.

Those scenarios have some common points. The three
scenarios include multi-scale physics in which a broad range of length and time scales play an 
important role. In the supernova core collapse case the length scales to cover range from the size
of the iron core ($\sim 1000$~km) to the size of the PNS ($\sim 10$~km). The smaller length-scale
process dominating the dynamics ranges from $\sim 1$~km in the case of convectively unstable
PNS formed after the collapse of the most common slow rotating progenitors
(M\"uller \& Janka 1997), to $1$~cm - $1$~m if the magneto-rotational
instability develops in the collapse of fast rotating progenitors (Cerd\'a-Dur\'an \etal 2007).
Furthermore the mechanism for the explosion requires an appropriate modeling
of the neutrino transport. The energy deposited by neutrinos behind the
shock and the standing accretion-shock instability (SASI) probably are crucial
for a successful explosion (Marek \& Janka 2009). In the collapsar scenario
strong rotation is necessary for the formation of a GRB. It allows the
generation of an accretion disk around the formed black hole with a
low-density funnel along the rotation axis. The relativistic jet which
is responsible for the GRB is powered either by MHD processes or by the energy
deposition of annihilating neutrinos around the axis (see e.g. MacFadyen \etal 2001).
In the first case, magneto-rotational instabilities and dynamo processes are
most probably responsible for the amplification of the magnetic field during
the collapse. The case of dynamo processes in rapidly rotating PNS resembles the
modeling of the sun. The cooling time-scale is of the order of seconds while
the rotation period which drives the dynamo processes can be as low as $1$~ms
reaching magnetic Reynolds number of $\sim 10^{17}$ (Thompson \& Duncan 1993).
The numerical modeling of all scenarios described above require very high
resolution to resolve the MRI and the turbulence amplifying the magnetic field,
full three-dimensional simulations to correctly capture the growth and saturation
of the different types of instabilities (MRI, SASI, global MHD instabilities),
and a time evolution much larger than the characteristic dynamical timescale
of the system. 
Therefore the numerical  methods intended to solve this  problem should be as
less  dissipative  as  possible  in  order to  guarantee  that  the  numerical
dissipation is smaller  than the expected physical one or,  in cases where the
computational requirements  do not  allow this, the  global properties  of the
system  show  convergence  with  increasing  grid  resolution.  The  numerical
dissipation  could be  a serious  limitation  for the  successful modeling  of
objects mentioned above.
It is hence important to have the appropriate 
scaling with  the resolution, which is a necessary
property to have scalable numerical codes which can run efficiently in massive 
parallel computers.

Most of the numerical codes performing simulations of these scenarios use
Eulerian grids, explicit numerical schemes, and a mesh adapted to the
problem, mostly with grids in spherical polar coordinates. Eulerian grid-based codes 
are better suited for these scenarios than Lagrangian methods (e.g. smoothed particle 
hydrodynamics) because they allow to use finite-volume conservative schemes.
Eulerian methods allow for the correct treatment of arbitrarily high discontinuities and
shocks in general relativity (Iba\~nez \etal 2000, Dimmelmeier \etal 2002,
Duez \etal 2003, Shibata 2007, Baiotti \etal 2007), even with magnetic fields
(Gammie \etal 2003, Komissarov \etal 2005, Anninos \etal 2005, Ant\'on \etal 2006). Explicit
methods are better suited for multidimensional simulations since they are
computationally less expensive and easier to parallelize, although they have
time-step limitations given by the CFL condition (Courant \etal 1928, 1967).
Spherical polar coordinates have several properties which make them appropriate
to model the objects described above: i) are well adapted for quasi-spherical
objects, ii) allow for accurate conservation of angular momentum, which is not
true, in general, for Eulerian grids (see e.g. Zink \etal 2008 and Fragile \etal 2009), 
iii) axisymmetry and spherical symmetry
can be easily enforced, and iv) large radial domains can be covered using non
equally spaced radial grids. These numerical methods have been successfully
applied in 1D (spherical symmetry) and 2D (axisymmetry) simulations. 
However the extension to 3D in spherical coordinates suffers from severe time-step
restrictions which render simulations unaffordable unless a special treatment
of the central and polar regions is used (see. e.g Müller \& Janka 1997)

Numerical dissipation effects are present in simulations using Eulerian grids
due to the discretization of the equations that are solved. There are several
methods to quantify the amount of numerical dissipation of a code based on the
measure of the energy losses of the system. This
has been standard practice since the first studies of hydrodynamic turbulence
(e.g. Herring \etal 1974) due to the necessity of resolving the physical
dissipation scales. There are also studies of the decay of waves in
hydrodynamics (Porter \etal 1994) and MHD simulations (e.g. Simon \& Hawley
2009). More recently the numerical dissipation has been estimated measuring
the angular momentum transport by MHD turbulence (Fromang \& Papaloizou 2007, 
Simon \etal 2009). All these methods allow for a simplified numerical setup in
where the local dissipation properties of the numerical algorithms can
accurately be estimated. We propose an alternative approach to 
measure the numerical dissipation that is suitable for global simulations of
relativistic stars close to the equilibrium.

The aim of this paper is to study the effects of numerical viscosity in
simulations with spherical coordinates and study the influence of the grid
resolution and the numerical scheme. In
\sref{sec:grvisc} we describe the hydrodynamics equations
including physical bulk viscosity in general relativity (GR), in \sref{sec:visc}
we present a method to estimate numerical dissipation effects in a simplified
test case. We apply this method in \sref{sec:ns} to estimate the numerical
viscosity of oscillating neutron stars. We finish the paper in
\sref{sec:discussion} discussing the implications of our numerical results.
If is not explicitly mentioned, we use units in which $c=G=1$. Greek indices 
run from $0$ to $3$ and Latin indices from $1$ to $3$.

\section{GR hydrodynamics with bulk viscosity}
\label{sec:grvisc}

We use the 3+1 decomposition of the spacetime in which the metric reads
\begin{equation}
  ds^2 = g_{\mu\nu} \rmd x^{\mu} \rmd x^{\nu} = - \alpha^2 \, dt^2 + \gamma_{ij} (dx^i + \beta^i \,dt)
  (dx^j + \beta^j \, dt),
\end{equation}
where $\alpha$, $\beta^i$ and $\gamma^{ij}$ are the lapse function, the shift
vector and the spatial 3-metric respectively. In addition we consider the
conformally flat condition (CFC) approximation (Isenberg 2008, Wilson \etal
1996) for the 3-metric $\gamma_{ij} = \phi^4 f_{ij}$, where $\phi$ is the conformal factor and 
$f_{ij}$ the flat 3-metric in spherical coordinates. This approximation uses 
the maximal slicing condition and quasi-isotropic coordinates as gauge conditions.
Under this approximation
the resulting system consist in a hierarchy of elliptic equations 
(Cordero-Carri\'on \etal 2009).

To be able to study numerical dissipation effects we need a physical counterpart
to calibrate our results. We use for this purpose the bulk viscosity of the fluid.
The energy momentum tensor of a fluid with bulk viscosity (Ehlers 1961) is
\begin{equation}
T^{\mu\nu}=\rho(1+\epsilon) u^\mu u^\nu + (P - \zeta \Theta)h^{\mu\nu},
\end{equation}
where $\rho$ is the rest-mass density, $\epsilon$ is the specific internal
energy, $u^{\mu}$ is the 4-velocity of the fluid, $P$ is the pressure, $\zeta$
is the bulk viscosity, $\Theta \equiv u^\mu_{;\mu}$ is the expansion of the fluid
and $h^{\mu\nu}\equiv g^{\mu\nu} + u^{\mu} u^{\nu}$. 
The bulk viscosity appears as an isotropic term in the energy-momentum tensor
in a very similar way to the pressure. Therefore we can define a modified
pressure $\hat{P} \equiv P-\zeta \Theta$ such that the energy-momentum tensor
has the same form as a perfect fluid
\begin{equation}
T^{\mu\nu}=\rho \hat{h} u^\mu u^\nu + \hat{P} g^{\mu\nu},
\end{equation}
where $\hat{h} \equiv h- \zeta \Theta / \rho$, being $h \equiv 1+\epsilon + P/\rho$ the
relativistic specific enthalpy. Using this form of the energy-momentum tensor is easy to modify an
existing hydrodynamics code to include bulk viscosity, specially if the next
assumptions are taken into account: i) we approximate the expansion
assuming a post-Newtonian expansion and small perturbations, 
\begin{equation}
  \Theta = \bi{\nabla}\cdot \bi{u} + \mathcal O (c^{-2}) + \mathcal O (v^2),
  \label{eq:expansion_approx}
\end{equation}
where $\mathcal O (c^{-2})$ corresponds to first post-Newtonian corrections.
All the simulations that we run with physical bulk viscosity in this work belong to this
regimen. ii) The bulk viscosity coefficient $\zeta$ is
sufficiently small, such that $\hat{P}>0$. And iii) we neglect the contribution
of the bulk viscosity in the computation of the sound speed needed for our
numerical scheme.

The hydrodynamics equations with bulk viscosity can be cast as a system of
conservation laws (cf. Iba\~nez \etal 2000)
\begin{eqnarray}
\fl \partial_t \left( \sqrt{\gamma} D \right )
+ \partial_i \left[ \sqrt{\gamma} D v^{* i} \right] &=& 0, \label{eq:continuity}\\
\fl \partial_t \left ( \sqrt{\gamma} S_j \right ) 
+ \partial_i \left [ \sqrt{\gamma}\left( S_j v^{*i} + \delta^i_j \alpha
 \hat{P} \right) \right ] &=& \frac{1}{2}\alpha\sqrt{\gamma}T^{\mu\nu} \,\partial_j g_{\mu\nu}
, \label{eq:linear_momentum} \\  
\fl \partial_t \left ( \sqrt{\gamma} \tau \right ) 
+ \partial_i \left [ \sqrt{\gamma} \left [ \tau v^{*i}
+ \alpha \hat{P} v^i\right) \right ]  &=& \alpha^2 \sqrt{\gamma} \left ( 
T^{\mu 0} \partial_\mu(\ln{\alpha}) - T^{\mu\nu}\Gamma^0_{\mu\nu}
\right ) \label{eq:energy_eq}
\end{eqnarray}
where $D\equiv \rho W$, $S_j \equiv \rho \hat{h} W^2 v_j$,  $\tau \equiv \rho \hat{h} W^2 -
\hat{P} - D$ are the conserved variables, $v^{*i}\equiv \rmd x^i / \rmd t$
is the coordinate 3-velocity, $v^i = (v^{*i}-\beta^i )/\alpha$
is the 3-velocity as measured by an Eulerian observer, and
$W=1/\sqrt{1-\gamma_{ij}v^iv^j}$ the Lorentz factor.
In the non-relativistic limit the equations for a classical viscous fluid can be
recovered (Landau \& Lifschitz 1987) which for constant $\zeta$ result in the Navier-Stokes equations.

We solve the coupled system of CFC spacetime evolution and GR hydrodynamics equations
using the numerical code \texttt{COCONUT} (Dimmelmeier \etal 2002, 2005). The
numerical code uses standard high-resolution shock-capturing schemes for the
hydrodynamics evolution in spherical polar coordinates, and spectral methods 
for the spacetime evolution.

\section{Estimating numerical viscosity}
\label{sec:visc}

\begin{figure}
  \centering
  \resizebox{0.48\textwidth}{!}{
  \includegraphics*{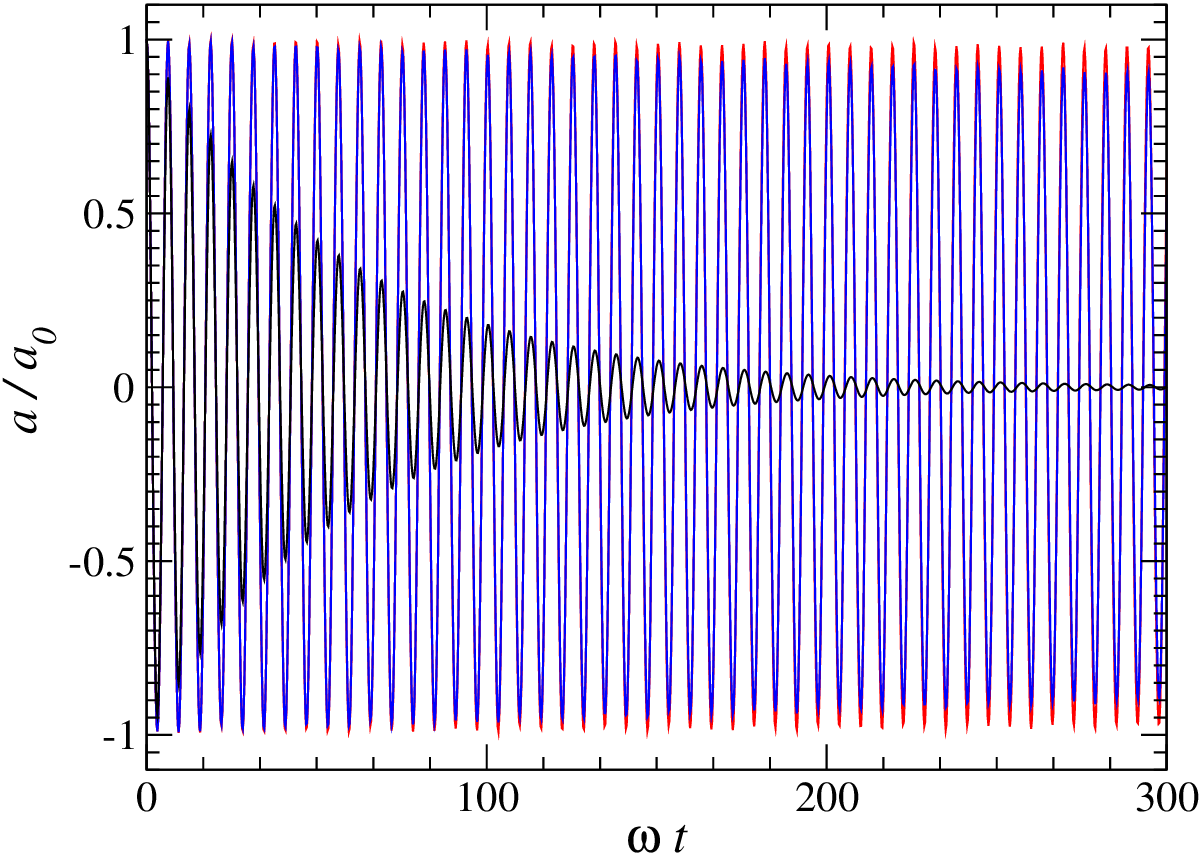}}
  \hspace{0.02\textwidth}
  \resizebox{0.48\textwidth}{!}{
  \includegraphics*{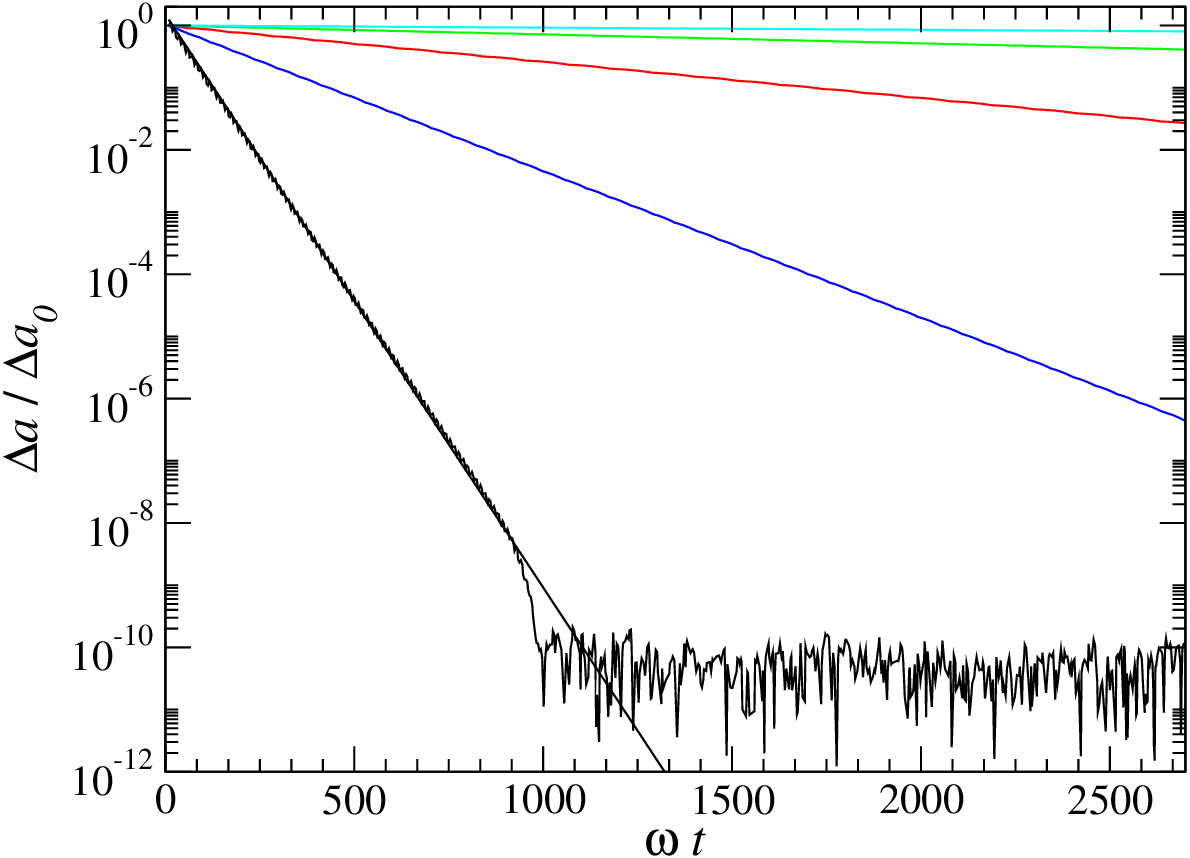}}
  \caption{Time evolution of the first
    oscillation mode in the spherical test system. All evolutions shown
  are computed using the Marquina flux formula. Left panel shows the evolution
  of the perturbation $a$ normalized to its initial value $a_0$ for a radial
  resolution $n_r=80$ and different reconstruction
  procedures: constant (black line),  minmod 
  (blue line) and MC  (red line).
  The right panel shows the evolution of the amplitude of the perturbation $\Delta a$ 
  normalized to its initial value $\Delta a_0$ using minmod
  reconstruction  and different resolutions:
  $10$ (black), $20$ (blue), $40$ (red), $80$ (green) and $160$ (cyan).
  We also plot the fit of the lowest resolution simulation to an exponential
  decay (black line).}
  \label{fig:st_time_evol}
\end{figure}
To estimate the numerical viscosity of our code we have designed
a simple spherical test. We consider a spherical fluid system of radius $R$ and constant density in a 
static Minkowski spacetime. This system allow for discrete modes of radial
oscillations (see appendix A). Since the
eigenfunctions of the modes are known it is possible to excite very accurately
single modes of the system and follow their evolution. We evolve the system
using a polytropic equation of state, therefore, since the system is
adiabatic, there are no possible energy losses and the oscillations should 
keep constant amplitude as long as non-linear effects does not appear. Hence, any
damping observed in the numerical simulation has to be caused by numerical
dissipation effects.

We  have chosen a  system with  $R=1$ and  initial density  $\rho_0 =  1$. The
equation  of  state  is  a  polytrope  of  the  form  $P=K  \rho^\Gamma$  with
$\Gamma=4/3$  and $K=1/3\times10^{-3}$. We  use a  perturbation $A(r)$ (see
appendix A)  of the
velocity  with an amplitude  of $v'_0  = 10^{-5}$  corresponding to  the lowest
frequency  mode, $\omega  = 0.094$.  This amplitude  is sufficiently small   for the
oscillations to  be considered linear.  We have used second  order Runge-Kutta
method for  the time evolution in  all cases and two  different flux formulae,
Marquina  (Donat el al 1998)  and HLL  (Harten \& van Leer 1983).  We have  computed all  models using
different reconstruction  techniques (constant and linear with minmod or MC slope limiters) 
and  different resolutions ($n_r = 10$, $20$,  $40$, $80$ and $160$).
We have  also computed the models without  physical bulk viscosity ($\zeta =  0$) and with
non-zero values  ($\zeta=10^{-4}$, $10^{-5}$ and $10^{-6}$). We  have evolved the
system  for  $2700 \,\omega^{-1}$. To follow the evolution of the oscillations we computed the quantity
\begin{equation}
a (t) = \int \rmd x^3 A(r) v^r(r,t).
\end{equation}

\begin{figure}
  \centering
  \resizebox{0.48\textwidth}{!}{
  \includegraphics*{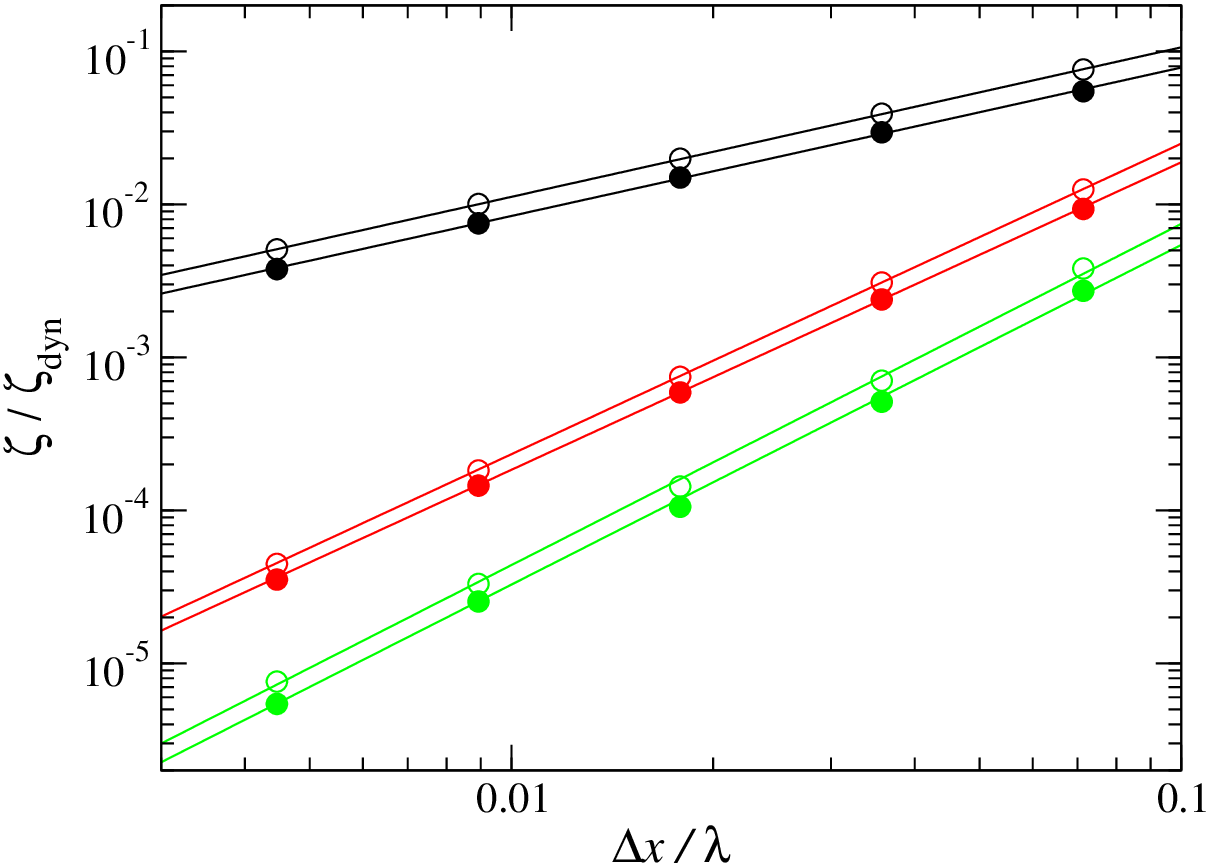}}
  \hspace{0.02\textwidth}
  \resizebox{0.48\textwidth}{!}{
  \includegraphics*{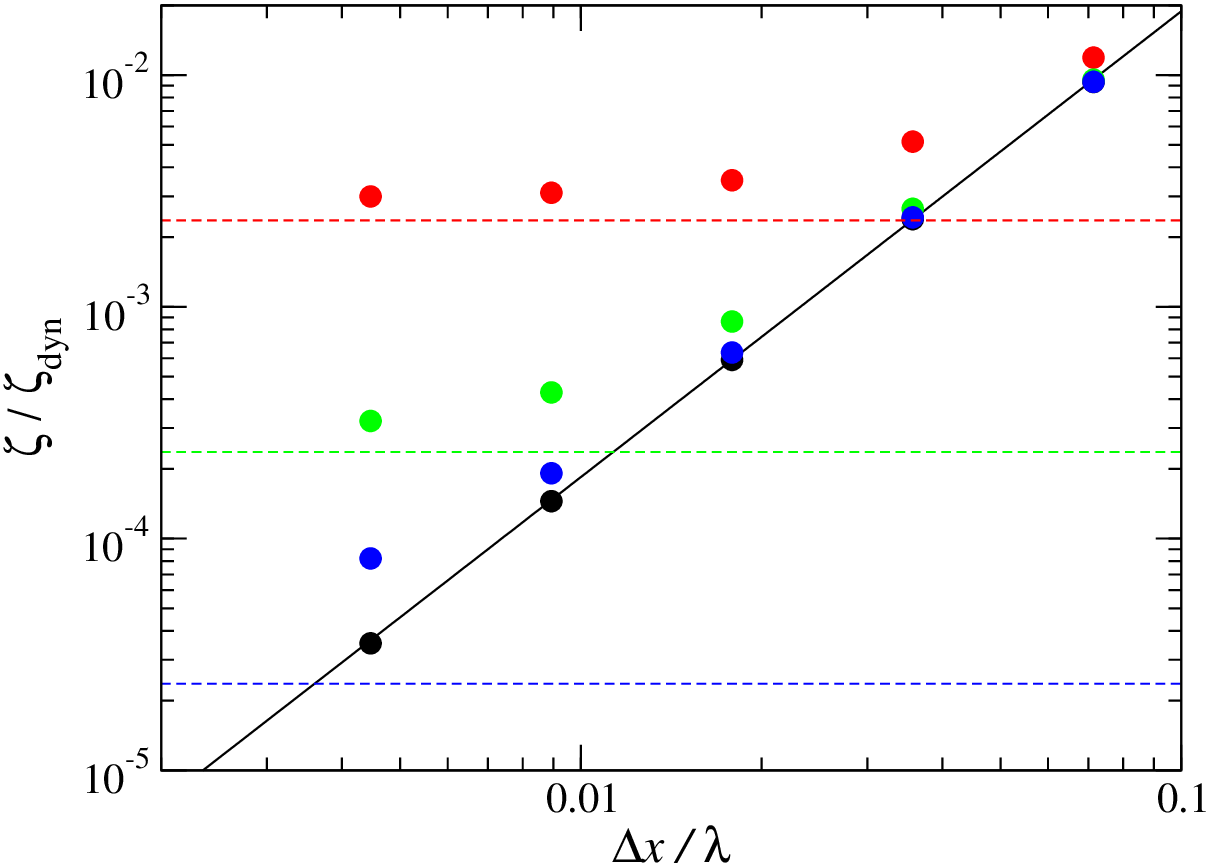}}
  \caption{Dependence of the numerical viscosity $\zeta$ with the resolution in the case of the test sphere. 
  The left panel shows the case without physical bulk viscosity for different flux formulae, 
  Marquina (filled circles) and HLL (open circles), and different reconstruction schemes,
  constant (back), minmod (red) and MC (green). The right panel shows the case with Marquina
  flux formula and MC reconstruction for different values of the physical viscosity $\zeta= 10^{-4}$ (red)
   $10^{-5}$ (green), $10^{-6}$ (blue) and $0$ (black). Dashed lines of the same color correspond to the
   value of the physical viscosity in each case.}
  \label{fig:visc_scaling}
\end{figure}

Left panel of  \fref{fig:st_time_evol} shows the time evolution of $a/a_0$ for
different reconstruction procedures, $a_0$ being the initial value of $a$.  The oscillation frequency coincides with
the predicted by  the linear analysis within $0.1\%$ for $n_r=80$ . It
can be seen that the order of the reconstruction has a strong influence in the
numerical  damping  of  the  system,  being  much  stronger  for  first  order
reconstruction (constant) than  for second order (minmod or MC).  In order to
quantify this effect we estimate the damping time for each simulation as it is described 
next. In every  case, we  first compute  the amplitude  of  $a$ as  the difference  between
two consecutive  oscillations, $\Delta a$. In  the right  panel of  \fref{fig:st_time_evol} we
plot the  evolution of the amplitude  for the minmod reconstruction case for five
different  resolutions.  For  very   low  resolution,  the  amplitude  of  the
perturbation  falls bellow the  round-off error  of the  code within the duration
of  the simulation and  it saturates. We fit next the
the  amplitude  of the oscillations to   an  exponential  decay  $\Delta a(t)  =
\Delta a (0)  \, \rme^{-2 t/\tau}$ for each model. This procedure allow us to compute
 the  value of the damping time of the
amplitude, $\tau/2$. Since  the energy of the perturbations  scales quadratically with the velocity, the
damping time of the energy is $\tau$. Finally, using \eref{eq:tau} it is possible to compute
the physical bulk viscosity from the value of $\tau$ as
\begin{equation}
\zeta= \frac{\zeta_{\rm dyn}}{\omega \, \tau},
\label{eq:visc_formula}
\end{equation}
where $\zeta_{\rm dyn}\equiv 2 \rho_0 c_s^2 / \omega$ and $c_s$ is the sound speed. 
The left panel of \fref{fig:visc_scaling} shows the behavior of the numerical viscosity with the
radial resolution for the different numerical schemes tested in simulations
without physical bulk viscosity ($\zeta=0$). To check the scaling with resolution we fit the viscosity values
to a power law
\begin{equation}
  \frac{\zeta}{\zeta_{\rm dyn}} =  S \left(\frac{\Delta x}{\lambda}\right)^p,
  \label{eq:visc_fit}
\end{equation}
where $\lambda = 2\pi c_{\rm s} \omega^{-1}$ is the wavelength of the oscillation mode, 
being $c_{\rm s}$ the speed of sound.
The results of the fitted parameters $S$ and $p$ for the different numerical methods are
shown in \tref{tab:fit}. In all cases $S$ is of order one, although in general the
Marquina flux formula provides about $30\%$ less bulk viscosity than the same
simulation with HLL. 
The scaling does not change substantially with the flux formula but
only with the reconstruction scheme: for constant reconstruction we recover
first order convergence and for linear reconstruction (minmod and MC) second order. These fits allow us
to compute the numerical bulk viscosity of different numerical schemes.

In the right panel of \fref{fig:visc_scaling}  we consider the simulations with 
physical bulk viscosity ($\zeta\ne 0$). In this case
the viscosity of the code decreases with resolution
with a similar scaling as in the case without physical viscosity. For
sufficiently high resolution, the viscosity converges to the value of the
added physical viscosity. Our method to estimate the bulk viscosity
overestimates, in the convergent regime, the real physical viscosity by $\sim25\%$.
We have checked that the damping time and therefore the viscosity is not affected
for a wide range of amplitudes of the perturbation ($0.1$-$10^{-7}$) and hence
we can conclude that non-linear effects does not play any role in the damping
observed during this test. We have also checked
that the results are independent of the CFL factor used for the time-step, by changing
its value between $0.8$ and $10^{-3}$. Since our final aim is to perform multidimensional
simulations we have also performed some axisymmetric 2D models of this
spherical system. We have chosen two representative radial resolutions
$n_r=20$ and $80$ and varied the angular resolution $n_\theta = 8$, $16$ and
$32$. In all cases the results are
indistinguishable from the 1D spherical simulations with the same number of
radial points.

\begin{table} 
\caption{\label{tab:fit} Coefficients $S$ and $p$ of the fitted numerical viscosity in radial oscillations
of a test fluid sphere, for different flux formulae and reconstruction schemes.} 
\begin{indented} 
\item[]
\begin{tabular}{@{}llllll} 
 \br Flux formula & reconstruc. & $p$ & $S$  \\
 \mr 
  Marq.& constant   &0.97&0.74 \\ 
  HLL & constant   &0.98&1.06 \\ 
  Marq.& minmod &2.01&1.91 \\ 
  HLL & minmod &2.03&2.66 \\ 
  Marq.& MC     &2.23&0.96 \\ 
  HLL & MC     &2.23&1.28 \\  
\br 
\end{tabular} 
\end{indented} 
\end{table}

\section{Numerical damping of neutron stars}
\label{sec:ns}

We apply our method to estimate viscosity in the case of radial oscillations
of non-rotating neutron stars. Our initial model is a non-rotating relativistic equilibrium 
configuration (Tolman 1939, Oppenheimer \& Volkoff 1939). We use a polytropic equation of state $P=K \rho^\Gamma$
with adiabatic index $\Gamma=2$  and polytropic constant $K=100$ (units of $G=c=M_{\odot}=1$).
We choose central rest mass density to be $\rho_{\rm c} = 7.9\times10^{14}$~g~cm$^{-3}$
and gravitational mass $M_{\rm g} = 1.4 M_{\odot}$. The  resulting
circumferential  radius is $R_{\rm C}=14.16$~km,  and the isotropic radial
coordinate at the surface of the star is $R=12.0$~km.
We evolve the system in dynamic spacetime adding an initial perturbation of the radial velocity 
corresponding to the fundamental radial oscillation mode with an amplitude of $10^{-3}$. 
To compute the eigenfunction of the fundamental mode necessary for the perturbation, we 
use the eigenfunction recycling technique described in Dimmelmeier \etal 2006.
We perform 1D spherical simulations with different radial resolutions of the
fluid grid, $n_r= 80$, $160$, $320$ and $640$,
which is equally spaced from the center to 14.4 km. 
The time-step is computed using the CFL condition for the eigenvalues of the 
hydrodynamics system, resulting in $\Delta t = 1.13\times 10^{-3}$~ms for a
grid with $n_r=80$ and a CFL factor 0.8.
We use an artificial atmosphere to treat the vacuum surrounding the star
as described in Dimmelmeier \etal 2002. The treatment consists of resetting 
all numerical cells with $\rho$ bellow a certain threshold to the value
$\rho_{\rm atm}$ and setting the velocity to zero. The values for the threshold
and the atmosphere are $10^{-6}$ and $10^{-10}$ times the initial central density 
respectively. We have checked that the results presented here do not change
if we decrease the values of these two quantities.
 We use three radial domains for the spectral metric solver: two including the
star and one compactified domain in the exterior. Each of the domains has either $17$ or $33$
collocation points for our low and high metric resolution simulations. Since the CFC approximation
only contains elliptic equations, which are not
restricted by the CFL condition, it is not necessary to solve the metric 
as frequently as the hydrodynamics equations.
We use a metric computation rate of $n_r/8$, i.e. we compute the metric once every $n_r/8$ 
time step of the hydrodynamics. We use a parabolic extrapolation of all the metric quantities between
consecutive metric computations, which has been shown to provide sufficient accuracy during the evolution
(Dimmelmeier \etal 2002). 
We do not add physical bulk viscosity in any of the simulations ($\zeta=0$), therefore the only dissipative
processes in the evolution are of numerical nature.
To compute the numerical bulk viscosity we use \eref{eq:visc_formula} as in the previous section
but with $\zeta_{\rm dyn} \equiv 2 {\bar \rho}_0 c_s^2 / \omega$ according to \eref{eq:tau_ns},
where $\bar \rho$ is estimated as the central density.
Since the numerical viscosity can depend on the location in the star the resulting value 
represents an average numerical viscosity. For our
particular case and the fundamental radial mode ($\omega=9.11\times 10^3$~Hz) the 
resulting value is $\zeta_{\rm dyn}=3.99\times10^{31}$~g~cm$^{-1}$~s$^{-1}$. 
This procedure gives us an idea of the order of magnitude of the averaged numerical viscosity on the star.
More accurate computations of the damping-time could be done using the expressions of the appendix B or 
the procedure described in Cutler \etal 1990.

\begin{figure}
  \centering
  \resizebox{0.48\textwidth}{!}{
  \includegraphics*{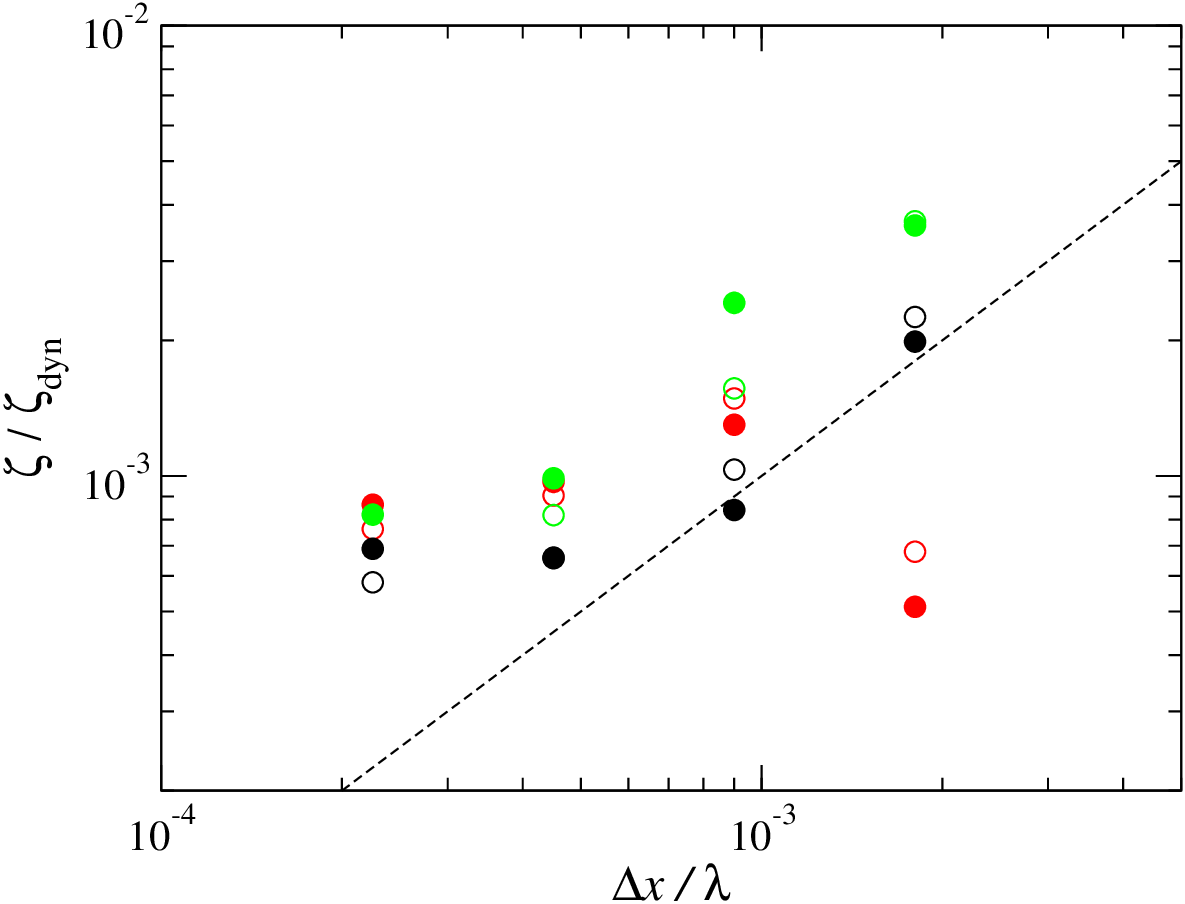}}
  \caption{Dependence of the numerical viscosity $\zeta$ on the resolution in neutron star evolution.
    We show values for different flux formulae, 
  Marquina (filled circles) and HLL (open circles), and different reconstruction schemes,
  constant (back), minmod (red) and MC (green). In all cases the highest metric resolution case is plotted.
  The dashed line represents the function $\zeta / \zeta_{\rm dyn} = \Delta r / \lambda$.
  }
  \label{fig:visc_scaling_rns}
\end{figure}

\Fref{fig:visc_scaling_rns} shows the variation of the viscosity with the radial resolution
normalized to the wavelength of the perturbation computed at the center of the star.
In general the viscosity decreases with increasing resolution. However the lowest resolution case
with the minmod reconstruction shows an anomalously low numerical viscosity, which we think is an artifact
of the low resolution. We plot the function $\zeta / \zeta_{\rm dyn} = \Delta r / \lambda$ for comparison.
The general trend is roughly first order decreasing the order for the highest resolution models,
although we are using  both first and second reconstruction schemes. 
The reason for the lower order of convergence is that all reconstruction schemes reduce to
first order  at  discontinuities.  Since we encounter a discontinuity at the
surface any evolution will inevitably lead to first order convergent results.
This is also a strong indication that the main numerical 
dissipation process in the neutron star evolution is probably due to the surface of the star, and therefore
increasing the reconstruction order does not decreases the mean numerical viscosity responsible for the damping
of global oscillations.

We have checked the influence of the CFL factor, the metric calculation rate and  the 2D effects for simulations
with $n_r=80$ radial points, Marquina flux formula and MC reconstruction. 
One of the main differences with respect
to the spherical test case of the previous section is that the equations that we are solving have sources due to the
gravity terms. In general the presence of non-zero sources can increase the stiffness of the problem leading to
inaccurate solutions, even unstable for extremely stiff sources. We expect
some influence of the CFL factor in our simulations since the reduction of the time step tends to cure the stiffness problems. 
We have performed simulations with CFL factor $0.8$ (standard for all
previous simulations), $0.1$, $0.01$ and $0.001$, with a metric computation
rate of  $10\times 0.8/{\rm (CFL \,factor)}$, which maintain 
constant the metric computation rate per unit evolution time. The amplitude of the oscillation at the end of
the simulation varies about $0.5\%$ at most. However most of the variation is
observed between CFL factor $0.8$ and $0.1$, and  
decreasing the CFL factor further does not significantly changes
the solution. Therefore the effect on the computed damping rate can be neglected.
We find that the stiffness of the sources depends on how often the metric is computed. We also perform the same 
simulations with a metric computation rate $10$. In this case the metric
is computed more often per time unit than the previous one, as the CFL factor decreases.
We find variations of about $25\%$ in the final amplitude depending on the CFL
factor used,  
although these variations depend strongly on the radial resolution. For $n_r=160$ the maximum variation
due to CFL factor changes is smaller than $10\%$. Since the effect of the CFL factor becomes larger lowering the
metric computation rate, we can conclude that the sources
become more stiff if the metric is computed more often.  
An explanation for this effect is  that if the metric is computed less often,
the sources vary  due to the space-time evolution  on time-scales longer, than
the hydrodynamic variables, which reduces the stiffness.
We conclude that the stiffness of the sources is not significantly affecting the computation
of the damping rates for the metric computation rate that we use in our regular simulations. However special care should
be taken in simulations in which the metric computation rate is high (close to one), e.g. in black hole formation simulations,
since the stiffness of the sources could lead to an increasing of the numerical viscosity of the code if the CFL factor
is not appropriately lowered. We also perform 2D-axisymmetric simulations with $4$, $8$
$16$ and $32$ angular grid points. Since the initial perturbation is radial, non-radial modes are not excited. We find 
an error in the angular velocity which is of order  $v^\theta/v^r \approx 10^{-9}$, and which is closely related to 
the accuracy of numerical recovery of the primitive variables. Since the angular extension of the grid restricts the 
time step of the simulation in a factor proportional to $1/n_\theta$, we find a similar behaviour increasing $n_\theta$
as it is found decreasing the CFL factor.

\section{Discussion}
\label{sec:discussion}

We conclude that it is possible to estimate dissipation effects in numerical simulations which produce
an effect similar to physical bulk viscosity. In a very simplified case,
as the test fluid sphere of constant density, the numerical viscosity scales with the order of the
reconstruction. In the application of our method to neutron stars simulations we find that
the behaviour of the numerical viscosity is more complicated since the gravity sources also 
contribute to the numerical dissipation processes, and the presence of the surface lowers the scaling 
of the viscosity to first order.

To understand why the numerical inaccuracies can be modelled as a bulk viscosity term we consider
the generic form of the flux formula in a Riemann solver
\begin{equation}
F^i_{\rm num} = \bar{F}^i + \Lambda \Delta U,
\end{equation}
where $F^i_{\rm num}$ is the numerical flux that is used in the time update, $\bar{F}^i$
is an averaged value of the flux,  $\Delta U$ represents the discontinuity at each cell
interface and $\Lambda$ the eigenvalue matrix. The particular form of the average and the matrix $U$
depends on the particular Riemann solver. If a reconstruction scheme of order $p$ is used,
the value of $U$ at both sides of the interface in smooth regions of the flow agrees up
to the $p-1$-th derivative, therefore for any function $f(U)$
\begin{equation}
 \Delta f (U) \propto \left (\Delta r \right )^p  {\mathcal D}^p U
\end{equation}
where ${\mathcal D}^p$ represents $p$-th spatial derivatives.
If one applies this expression to the momentum equation
\begin{equation}
\partial_t \left (\sqrt{\gamma } S_j \right) + \partial_i \bar {F}^i + \mu\, (\Delta r)^p \,    {\mathcal D}^{(p+1)} S_j = Q,
\label{eq:hydro_nvisc}
\end{equation}
where $\mu$ depends on the specific Riemann solver used and $Q$ represent the sources (r.h.s of
\eref{eq:linear_momentum}). The new term appearing in the numerical
version of the equations is a dissipative term since in includes a $p+1$ derivative, while
$\mu \, (\Delta r)^p$ is the corresponding dissipative coefficient. For first order reconstruction,
the dissipation term includes second derivatives that resemble to a viscosity term with viscosity
proportional to $\Delta r$. For higher order reconstruction the new term resembles hyperviscosity coefficients.
Note that with this interpretation we can identify the computed value of $\zeta$ in our simulations 
with a viscosity or hyperviscosity term, which scales with resolution with the order of the reconstruction 
scheme. Furthermore, we can argue that, since the new terms in \eref{eq:hydro_nvisc} are responsible
for both the bulk viscosity and shear viscosity terms, the value of the numerical shear viscosity
of the code has to be of the same order of magnitude as the numerical bulk viscosity estimated in this
work. If this case were true, the method described in this paper could be a powerful tool to 
estimate numerical dissipative effects in multidimensional simulations
.
However the method has some limitations due to the approximations that we
considered. The integral expressions \eref{eq:integrals_approx1} and
\eref{eq:integrals_approx2} as well as the approximate expression for the expansion
\eref{eq:expansion_approx} rely on the facts that a post-Newtonian expansion is
possible and that velocity perturbations are small. This provides 
an order-of-magnitude estimate in the case of systems
involving neutron stars or proto-neutron stars since $\mathcal O (c^{-2}) \sim
0.15$. In this scenarios the velocity involved in oscillations is typically smaller than
$0.1$. In the vicinity of black holes, the post-Newtonian expansion is not
convergent anymore, and the method can lead to large inaccuracies. Similar
thing happens in the case of flows with a high Lorentz factor as those observed
in jets.

From our simulations we also conclude that the gravity can be an important source of numerical viscosity 
which has to be estimated in an appropriate way. The stiffening of the sources in the presence of 
rapidly changing spacetime, can lead to a strong increasing in the numerical viscosity if the CFL
condition is not adapted accordingly. Although this effect is not a problem in the simulations presented
in this work, it could be in case in which the  spacetime evolves in similar time-scales as the
fluid, e.g. in the formation of a black hole. It is therefore important to test in the future which are the
real effects of numerical viscosity in such simulations.

Finally, we note that all the viscosity results given in this paper are normalized to $\zeta_{\rm dyn}$,
which depends on the frequency of the oscillation $\omega$. For a given astrophysical scenario,
and the same numerical resolution, different modes will be affected in a different way depending on the 
frequency of the mode. Therefore, in order to estimate the resolution needed to evolve a system for a given
amount of time with reasonably small damping one has to use the mode with higher frequency, of those 
who may be important in the dynamics of the system. Deciding which is this mode may be non trivial in
non-linear simulations, since a strong numerical damping in high frequency modes, which are coupled to 
the lower frequency modes, can modify the damping times for the low frequency modes too.

\ack 
This work was supported by the Collaborative Research Center on Gravitational Wave Astronomy 
of the Deutsche Forschungsgesellschaft (DFG SFB/Transregio 7). I would like to thank
Ewald M\"uller for his comments on the paper and Klaus Dolag for some useful discussions
about viscosity.

\appendix

\section{Radial oscillations of an homogeneus sphere.}

We consider  a spherically symmetric barotropic fluid  in Minkowski spacetime,
with   equation   of    state   $P=P(\rho)$.   The   hydrodynamics   equations
\eref{eq:continuity} and \eref{eq:linear_momentum} in this case read:
\begin{eqnarray}
\fl \partial_t \left( \rho W \right )
+ \frac{1}{r^2}\partial_r \left[ r^2 \rho W v \right] &=& 0, \label{eq:continuity}\\
\fl \partial_t \left ( \rho \hat h W^2 v \right ) 
+ \frac{1}{r^2}\partial_r \left [ r^2 \rho \hat h W^2 v^2 \right ] 
&=& -\partial_r \hat P
, \label{eq:linear_momentum}
\end{eqnarray}
where $v \equiv v^r$. The equilibrium solution of this system is trivially
zero velocity, $v_0=0$, and constant pressure, $P=P_0$, and hence constant rest mass density, $\rho=\rho_0$.
We consider perturbations of the rest mass density and velocity,
$\rho = \rho_0 + \rho'$ and $v = v'$, and hence
\begin{eqnarray}
P &=& P_0 +  \, \frac{\partial P_0}{ \partial \rho_0} \, \rho'
= P_0 + h_0 \, c_s^2 \, \rho'.
\end{eqnarray}
where $c_s^2$  is the speed of sound of the equilibrium model. 
The linearized hydrodynamics equations read
\begin{eqnarray}
\partial_t v' + \frac{c_s^2}{\rho_0} \partial_r \rho' 
- \frac{\zeta}{\rho_0 h_0} \partial_r \Theta &=& 0, \label{eq:vlinear}\\
\partial_t \rho' + \rho_0 \partial_r v' + \frac{2}{r} \rho_0 v' &=& 0, \label{eq:rholinear}
\end{eqnarray}
If  $\zeta$  is   sufficiently  small,  i.e.  $\zeta<<h_0\rho_0$,  the
viscosity  terms can  be neglected  in  the computation  of the  perturbations
spectrum. In this case equations \eref{eq:vlinear} and \eref{eq:rholinear} can
be combined in a wave equation for $\rho'$
\begin{equation}
\partial_{tt} \rho' - c_s^2 \partial_{rr} \rho' - \frac{2 c_s^2}{r} \partial_r
\rho' = 0.
\label{eq:rho_wave_eq}
\end{equation}
The system of equations  \eref{eq:vlinear} and \eref{eq:rholinear} admits 
oscillatory solutions of the form $v'=A(r)\, \sin{\omega t}$ and 
$\rho' = B(r)\,\cos{\omega t}$ and hence \eref{eq:rho_wave_eq} results in
\begin{eqnarray}
\partial_{rr} B + \frac{2}{r} \partial_r B
+ \frac{\omega^2}{c_s^2} B = 0.
\end{eqnarray}
This equation has the form of the Lane-Emden equation of index 1, which 
have solutions regular at $r=0$ (Chandrasekhar 1967) of the form 
\begin{equation}
  B (r) = \rho'_0 \frac{\sin{(k\,r)}}{k\, r},
\end{equation}
where $k = \omega / c_s$ and $\rho'_0$ is a parameter which
controls the amplitude of the density perturbation. From \eref{eq:vlinear}
the solution for the velocity perturbation is
\begin{equation}
  A (r) = v'_0 \, \frac{k\,r \cos{(k\,r) -\sin{(k\,r)}}}{(k\, r)^2},
\end{equation}
where $v'_0$ controls the amplitude of the velocity perturbation and is
related to $\rho'_0$ by
\begin{equation}
  \frac{\rho'_0}{\rho_0} = - \frac{v'_0}{c_s}.
\end{equation}
If we impose boundary conditions at the surface $v'(R)=0$, i.e., $A(R)=0$
we find a discrete spectrum of modes given by the solutions of
\begin{equation}
  \tan{(k\, R)} = k \, R.
  \label{eq:linearsol}
\end{equation}
We have computed the roots of this equations by means of a bisection
algorithm. The first five numerical solutions correspond to
$k\, R = 4.49$, $7.73$, $10.90$, $14.07$ and $17.11$. 

\section{Damping time of an oscillating spherical star}

The rate of change of the energy due to bulk viscosity of a pulsating star
is (cf. Cutler \etal 1990)
\begin{equation}
\frac{\rmd E}{\rmd t} = - 4 \pi \int_0^R \rmd r \, r^2 \phi^6 \,
\zeta|\Theta|^2 ,
\label{eq:e_loss}
\end{equation}
and, provided that the energy of the pulsations $E$ is known, the 
damping time is 
\begin{equation}
  \tau = -2E \left< \frac{\rmd E}{\rmd t} \right >^{-1}.
\label{eq:e_damping}
\end{equation}
where $<>$ denotes the time average over a cycle.

The energy stored in the radial oscillations of a spherical star can be 
computed as the energy difference between the star in equilibrium and the 
perturbed system. Previous computations (Meltzer \& Thorne 1966, Glass \& Lindblom 1983)
used Schwarzschild coordinates in their computations. Instead of performing
the coordinate transformation to the choice of the present work we find it easier 
and more instructive to compute the energy of the oscillations
directly in our coordinates, although the result should be identical.

The ADM energy is a conserved quantity which in spherical symmetry and 
under our gauge choice is
\begin{equation}
\fl E_{\rm ADM}= -2 \int_0^R \rmd r \, r^2 \Delta \phi 
= 4 \pi \int_0^R \rmd r \, r^2 \phi^5 \left (\rho h W^2 - P +
\frac{K_{ij}K^{ij}}{16 \pi} \right),
\end{equation}
being $\Delta$ the Laplacian with respect to the flat 3-metric and $K_{ij}$ the extrinsic 
curvature of the induced 3-metric $\gamma_{ij}$. Since the extrinsic curvature vanishes
for the equilibrium system in our gauge choice the equilibrium ADM energy is 
\begin{equation}
E_{\rm ADM \, 0}
= 4 \pi \int_0^R \rmd r \, r^2 \phi_0^5 (\rho_0 h_0 - P_0).
\end{equation}
The ADM energy does not change with time and hence we can compute its value by evaluating 
the integral at the oscillation phase with maximum velocity. 
In this phase, due to the continuity equation \eref{eq:continuity}, the variation
of $\phi^6 \rho W$ with respect to the equilibrium is zero. The leading term in the 
perturbation corresponds to quadratic terms in the velocity. The energy of the 
oscillations is thus $E=E_{\rm ADM} - E_{\rm ADM\, 0}$ which results in
\begin{eqnarray}
\fl E = && 
 4 \pi \int_0^R \rmd r \, r^2 \phi_0^5 
 \left [
\rho_0 h_0  \frac{1}{2} {v'}^2 
- ( \rho_0 h_0 - 5 P_0) \frac{\phi'}{\phi}  
+ \frac{{K'}_{ij}{K'}^{ij}}{16 \pi}
\right ]
\label{eq:e_pert}
\end{eqnarray}
for adiabatic perturbations.
Note that here $\phi' = \phi-\phi_0$ corresponds to the variation of $\phi$ with respect to the 
equilibrium for the phase in which $v'$ is maximum and hence is a term quadratic in $v'$.

If we apply this expressions to the case of the spherical fluid in  Minkowski spacetime of 
the appendix A the resulting expressions are
\begin{eqnarray}
E = 4 \pi \int_0^R  \rmd r \, r^2 \frac{1}{2} \rho_0 A(r)^2 =
\rho_0 {v'}^2_0 \pi \frac{R}{k^2} \sin^2{(k\, R)},
\\
\left < \frac{\rmd E}{\rmd t} \right > = 
- \zeta {v'}^2_0 \pi  R \sin^2{(k\, R)},
\end{eqnarray}
where we have explicitly used that \eref{eq:linearsol} is fulfilled to 
perform the integration
and that $<|\Theta|^2> = (\nabla A)^2 / 2 = \omega^2 B^2 / ( 2 \rho_0^2 )$.
The damping time is therefore
\begin{equation}
  \tau = \frac{2 \rho_0  c_s^2}{\zeta \omega^2},
\end{equation}
It is convenient to express it as
\begin{equation}
  \omega \tau = \frac {\zeta_{\rm dyn}}{\zeta}
  \label{eq:tau}
\end{equation}
where $\zeta_{\rm dyn} \equiv 2 \rho_0 c_s^2 / \omega$. For $\zeta \sim \zeta_{\rm dyn}$
the damping of the
mode occurs in dynamical time-scales, while if $\zeta << \zeta_{\rm dyn}$
the damping occurs in secular timescales.

In the case of a perturbed relativistic star the computation of the damping 
rate can not be computed analytically. However we can still make an order of 
magnitude estimation. For this purpose it is useful to truncate the equations
\eref{eq:e_loss} and \eref{eq:e_pert} to the leading order
in the post-Newtonian expansion
\begin{eqnarray}
E &=& 4 \pi \int_0^R \rmd r \, r^2 \frac{1}{2}  \rho_0  {v'}^2 
(1 + \mathcal O (c^{-2})),
\\
\frac{\rmd E}{\rmd t}& =& - 4 \pi \int_0^R \rmd r \, r^2  \, \zeta |\Theta|^2 
\left ( 1 + \mathcal O (c^{-2}) \right ).
\end{eqnarray}
which corresponds to the Newtonian expression for the kinetic energy, beging 
$\mathcal O (c^{-2})\sim \mathcal O (v^2) \sim \mathcal O (GM/R) $ the first post-Newtonian 
corrections (see e.g. Blanchet \etal 1980).
If we assume $e^{i (\omega t+ k r)}$ dependence of the 
perturbations, where $k$ is the wavenumber, then $<|\Theta|^2> \sim k^2
<v^2>=k^2 v_{\rm max}^2 / 2$. The energy and energy
losses result in this case
\begin{eqnarray}
E = 2 \pi \int_0^R  \rmd r \, r^2  \rho_0 v_{\rm max}^2
\left( 1 + \mathcal O (c^{-2})\right),
\label{eq:integrals_approx1}
\\
\left< \frac{\rmd E}{\rmd t} \right > 
\sim - 2 \pi k^2 \zeta \int_0^R  \rmd r \, r^2 v_{\rm max}^2 
\left( 1 +  \mathcal O (c^{-2})\right),
\label{eq:integrals_approx2}
\end{eqnarray}
and hence
\begin{equation}
 \omega \tau \sim \frac{2 \omega {\bar \rho}_0}{k^2 \zeta} =
\frac{2 {\bar \rho}_0  c_s^2}{\zeta \omega},
=\frac {\zeta_{\rm dyn}}{\zeta},
  \label{eq:tau_ns}
\end{equation}
where $\bar \rho$ is an averaged density weighed by the eigenfunction $A(r)$
and $\zeta_{\rm dyn} \equiv 2 {\bar \rho}_0 c_s^2 / \omega$.


\References

\item
  Anninos P, Fragile P C and Salmonson J D 2005 {\it Astrophys. J.} {\bf 635} 723

\item
  Ant\'on L \etal 2006 {\it A\&A} {\bf 637} 296
\item
  Baiotti L \etal 2005 {\it Phys Rev D} {\bf 71} 024035
\item
    Cerd\'a-Dur\'an P, Font J A , Ant\'on L and M\"uller E 2007 
    {\it Astron. Astrophys.} {\bf 492} 937
\item
  Cerd\'a-Dur\'an P, Font J A and Dimmelmeier H 2007 {\it Astron. Astrophys.} {\bf 474} 169
\item
  Chandrasekhar S 1967 {\it An introduction to the Study of Stellar Structure}
  {New York:Dover} p~84
\item
  Cutler C, Lee L and Splinter R J 1990 {\it Astrophys. J.} {\bf 363} 603
\item
  Courant R, Friedrichs K and Lewy H 1928 {\it Math. Ann.} {\bf 100} 32
\item
  Courant R, Friedrichs K and Lewy H 1967 {\it IBM J. Res. Dev.} {\bf 11} 215
  (English translation of Courant \etal 1928)

\item  
  Cordero-Carri\'on \etal 2009  {\it Phys. Rev. D} {\bf 79} 24017

\item
  Dimmelmeier H, Font J A and M\"uller E 2002 {\it Astron. Astrophys.} {\bf 388} 917
\item	
  Dimmelmeier H, Novak J, Font J A, Ib\'a\~nez J M and Müller E 2005
  {\it Physi. Rev. D} {\bf 71} 064023
\item 
  Dimmelmeier H Stergioulas N and Font J A 2006 {\it Mon. Not. R. Astron. Soc.}
  {\bf 368} 1609-1630
\item
  Duez M, Marronetti P, Shapiro S L and Baumgarte T W 2003 {\it Phys. Rev. D.} {\bf 67} 024004

\item
  Donat R, Font J A, Ib\'a\~nez J M and Marquina A 1998 {\it J. Comput. Phys.} {\bf 146} 58

\item
  Fragile P C, Lindner C C, Anninos P and Salmonson J D 2009 {\it Astrophys. J.} {\bf 691} 482

\item Fromang S and Papaloizou J 2007 {\it Astron. Astrophys.} {\bf 476} 1113

\item Gammie C F,  McKinney J C and Toth G 2003 {\it Astrophys. J.} {\bf 589} 444

\item
  Glass E N and Lindblom L 1983 {\it Astrophys. J. S.} {\bf 53} 93

\item
  Harten A, Lax P D and van Leer B 1983 {\it SIAM Rev.} {\bf 25} 35
\item
  Herring J R, Orszag S A, Kraichnan R H and Fox D G 1974 {\it J. Fluid Mech.} {\bf 66} {417}

\item
  Iba\~nez J M, Aloy M A, Font J A, Mart\'i J M, Miralles J A and Pons J A 2000
{\it Proc. Conf. on Godunov methods:
  theory and applications (Oxford)} ed E F Toro (Kluwer Academic/Plenum Publishers) p~485--496
\item
  Isenberg J A 2008 {\it Int. J. Mod. Phys. D} {\bf 17} 265
\item 
  Komissarov S S 2005 {Mon Not R Astron Soc} {\bf 359} 801

\item 
  Landau L D and Lifshitz E M 1987 {\it Fluid mechanics} (Amsterdam: Elsevier) p~44
\item
  Marek A and Janka H T 2009 {\it Astrophys. J.} {\bf 694} 664
\item
  MacFadyen A I, Woosley S E and Heger A 2001 {\it Astrophys. J.} {\bf 550} 410
\item
  Meltzer D W and Thorne K S 1966 {\it Astrophys. J.} {\bf 145} 514
\item 
  Modjaz M \etal 2008 {\it Astrophys. J.} {\bf 135} 1136
\item
  M\"uller E and Janka H T 1997 {\it Astrophys. J.} {\bf 317} 140
\item 
  Porter D H and Woodward H P 1994 {\it Astrophys. J. S.} {\bf 93} 309

\item 
  Simon J B and Hawley J F 2009 {\it Astrophys. J.} {\bf 707} 883
\item
  Simon J B, Hawley J F and Beckwith K 2009 {\it Astrophys. J.} {\bf 690} 974 

\item
  Tolman R C 1939 {\it Phys. Rev.} {\bf 55} 364
\item
  Oppenheimer J R and Volkoff G M 1939 {\it Phys. Rev.} {\bf 55} 374
\item
  Shibata M 2003 {\it Phys Rev D} {\bf 67} 024033

\item
  Tompson C and Duncan R C 1993 {\it Astrophys. J.} {\bf 408} 194

\item 
  Wilson J R, Mathews, G J and Marronetti P 1996 {\it Phys. Rev. D} {\bf 54} 1317

\item 
  Woosley S E 1993 {\it Astrophys. J.} {\bf 405} 273

\item
  Zink B, Schnetter E and Tiglio M 2008 {\it Phys Rev. D} {\bf 77} 103015

\endrefs

\end{document}